\documentclass[epj]{svjour}
\usepackage{graphics}
\usepackage{hyperref}
\hypersetup{colorlinks = true, allcolors = blue}
\usepackage[utf8]{inputenc}
\usepackage{authblk}
\usepackage{hepparticles}
\usepackage{hepunits}
\usepackage{hepnames}
\begin{document}
\title{Tracking and Vertex detectors at FCC-ee}
\author{Nicola Bacchetta\inst{1} \and Paula Collins\inst{2} \and Petra Riedler\inst{2}  
}                     
\offprints{}          
\institute{INFN-Padova, Padova, Italy \and CERN, EP Department, Geneva, Switzerland}
\date{Received: \today / Revised version: \today }
%
\abstract{
The combined vertexing and tracking performance of the innermost part of the FCC-ee experiments must deliver outstanding precision for measurement of the track momentum together with an impact parameter resolution exceeding by at least a factor five that typically achieved at LHC experiments. 
Furthermore, precision measurements require stability and fiducial accuracy  at a level which is unprecedented in collider experiments.
For the innermost vertex layers these goals translate into a target hit resolution of approximately 3 $\mu$m together with a material budget of around 0.2\% of a radiation length per layer.  Typically this performance might be provided by silicon-based tracking, together with a careful choice of a low-mass cooling technology, and a stable, low mass mechanical structure capable of providing measurements with a low enough systematic error to match the tremendous statistics expected, particularly for the run around the Z resonance.  At FCC-ee, the magnetic field will be limited to approximately 2\,T, in order to contain the vertical emittance at the Z pole, and a tracking volume up to relative large radius is needed.  The technological solution could be silicon or gaseous based tracking, in both cases with the focus on optimising the material budget, and particle identification capability would be an advantage.  Depending on the global design, an additional silicon tracking layer could be added at the outer radius of the tracker to provide a final precise point contributing to the momentum or possibly time of flight measurement.  Current developments in monolithic and hybrid silicon technology, as well as advanced gaseous tracking developments provide an encouraging road map towards the FCC-ee detector.  The current state of the art and potential extensions will be discussed and a generic call for technology which could have a significant impact on the performance of an FCC-ee tracking and vertexing detector is outlined. 
\PACS{
     {PACS-key}{discribing text of that key}   \and
     {PACS-key}{discribing text of that key}
     } 
} 
\maketitle
%




\section{Introduction: tracking requirements for FCC-ee}

The tracking volume which makes up the innermost part of any FCC-ee detector must be capable of delivering outstanding performance across the full acceptance, down to approximately 120 mrad, and full momentum range, typically with full efficiencies down to 300 MeV/c and 98\% or better for muons down to 100 MeV/c transverse momentum.  An overview of proposed detector layout and performance requirements can be found in the FCC-ee Conceptual Design Report (CDR)~\cite{Abada:2019lih} and in this issue~\cite{Blondel:2021ema}.  A driving factor in the design is that the magnetic field will be limited to approximately 2\,T, in order to contain the vertical emittance at the Z pole, and a tracking volume up to relatively large radius will therefore be needed. It must be engineered in a way which results in minimum material in front of the external detectors and a stable structure which is capable of providing measurements with a low enough systematic error to match the tremendous statistics expected, particularly for the Z pole running.  

\begin{figure}[htb]
    \centering
    \resizebox{0.6\textwidth}{!}{\includegraphics{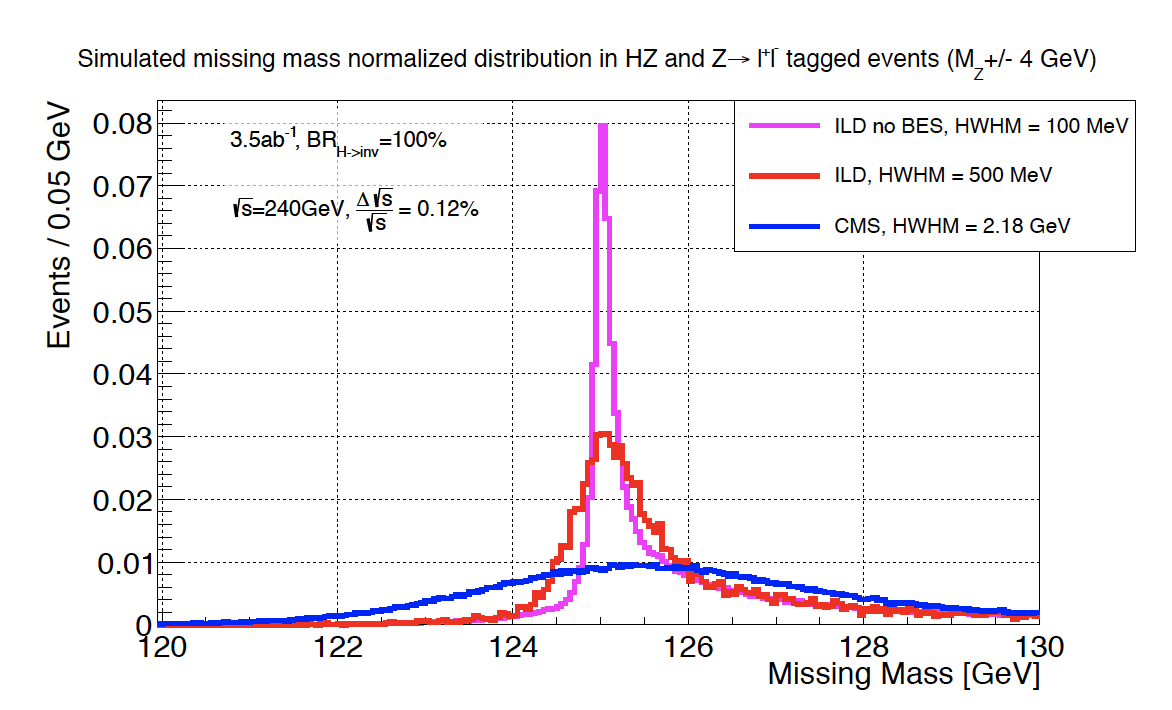}}
      \resizebox{0.35\textwidth}{!}{\includegraphics{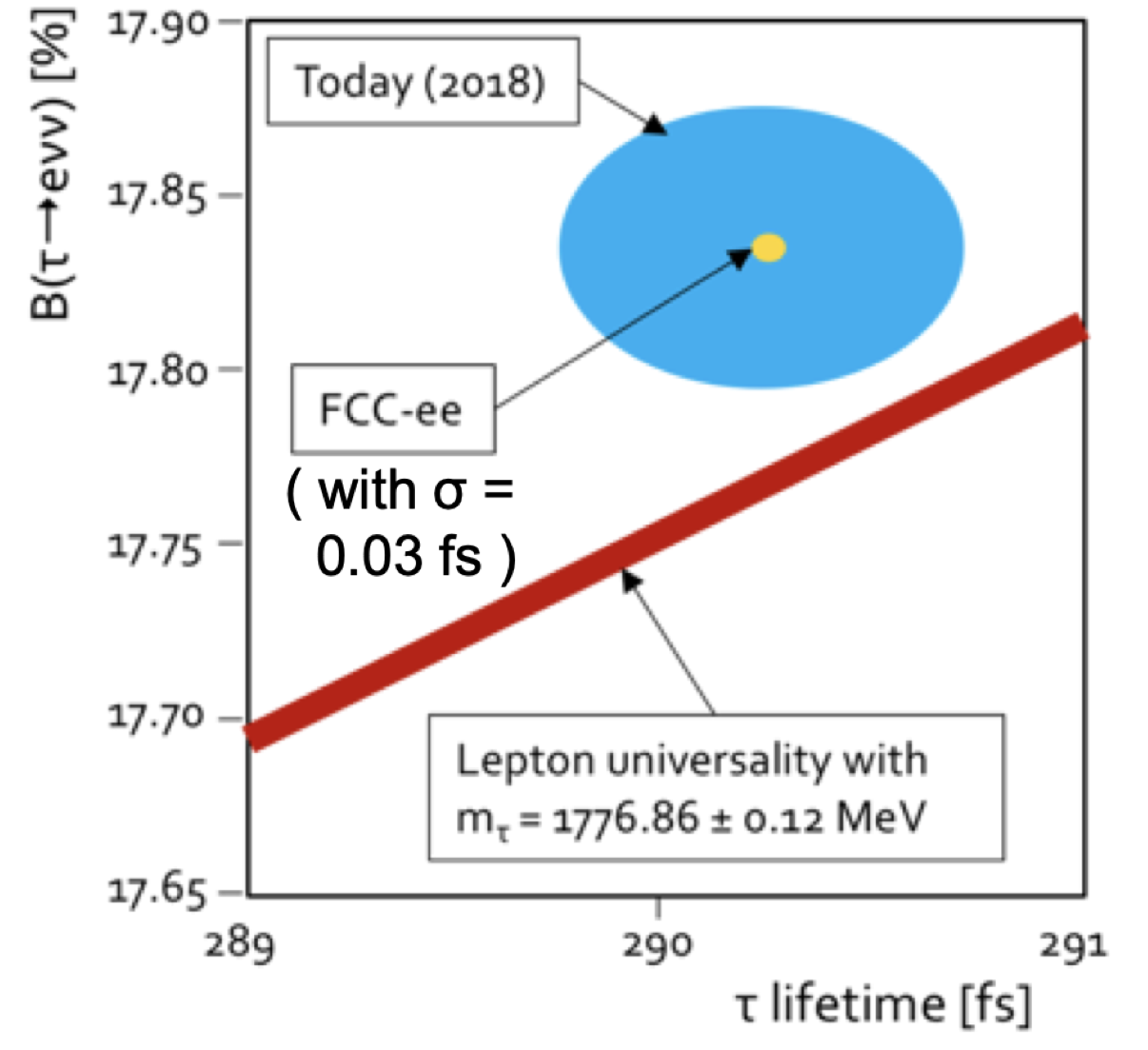}}
    \caption{Examples of measurements where the performance of the tracking system has a particular impact.  The left plot shows the missing mass distribution in $HZ$ and $Z \rightarrow l^+l^-$ events~\cite{Cerri:2016bew}, illustrating the difference in resolution between an ILD-like detector (red line) and a CMS-like detector (blue line), where a main cause of the improvement is the superior tracking resolution of the ILD-like reconstruction.  The right plot shows the potential access to sensitivity to the $\tau \rightarrow e \overline{\nu} \nu$ lepton universality test~\cite{10.21468/Dam:tau}, with the improvement in the $\tau$ branching fraction and lifetime measurements which can be expected at FCC-ee.  The tracking system contributes significantly to this expected improvement, in particular with the space point precision, low radius, and low material expected in the vertex detector.}
    \label{fig:ildmassres}
\end{figure}

\begin{figure}[htb]
    \centering
    \resizebox{0.9\textwidth}{!}{\includegraphics{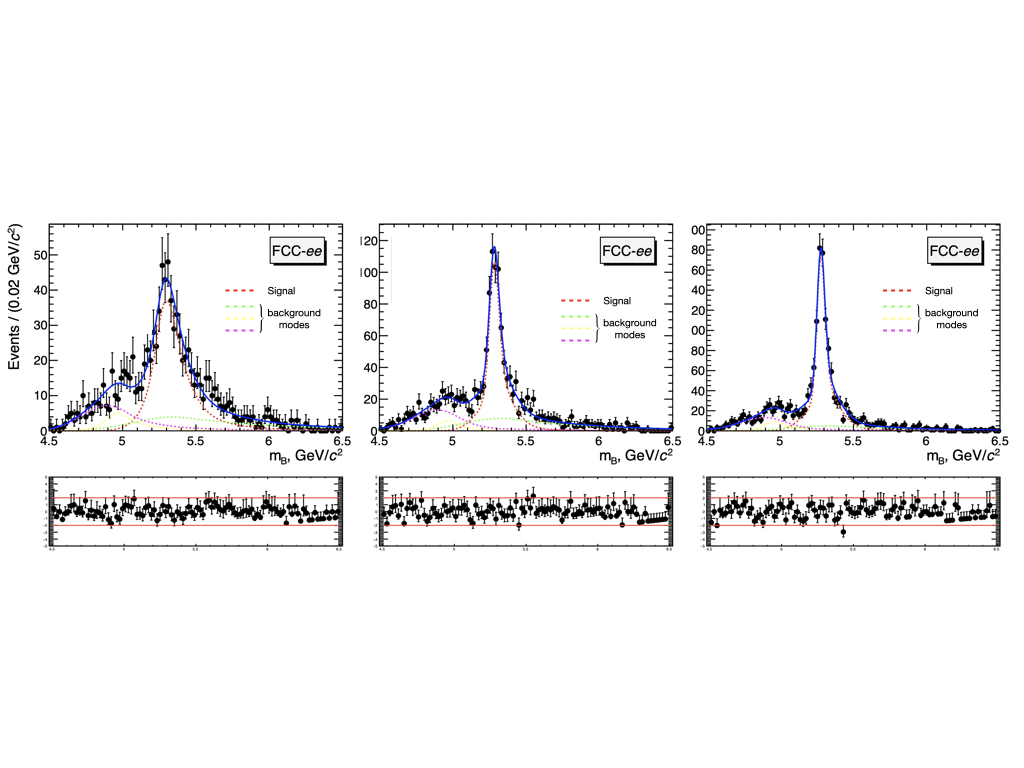}}
    \caption{The importance of the secondary vertex resolution in driving the signal to background quality is demonstrated in this study of the topologically reconstructed invariant mass of $B^o_d \rightarrow K^* \tau \tau$ candidates.  The leftmost plot shows the expected signal for a vertex detector with an ILD like performance.  The middle and right plots show the improvement that can be gained by artificially improving the vertex resolution of the vertex detector by a factor 2 (middle) and 4 (right)~\cite{stephane_gdr}.}
    \label{fig:leptonuniversality}
\end{figure}

The role of the tracking system will be decisive for the FCC-ee physics goals.  Examples of physics channels which place particular demands on the vertexing and tracking are listed here:

\begin{itemize}
    \item{The clean environment in which the Higgs will be produced at FCC-ee, with 1M ZH events expected at 240\,GeV~\cite{blas}, gives a unique opportunity to explore all decay modes, hence the importance of excellent b, c and $\tau$ tagging, placing high demands on the quality of the impact parameter resolution and secondary vertex resolution.}
    \item{Another unique feature of the clean Higgs production via ‘‘Higgsstrahlung" is the possibility of reconstructing the Higgs recoil mass against the Z boson~\cite{blas}.  In order to fully benefit from this, an excellent momentum resolution is required from the main tracker.}
    \item{Accessing $\tau$ properties at the so called ‘‘TeraZ", referring to the expected datasample of $5 \times 10^{12}$ collected Z bosons,  will be another major opportunity at FCC-ee~\cite{10.21468/Dam:tau}.  Measurements which require excellent vertexing include the lifetime, mass, leptonic branching fraction and lepton flavour violating decays such as $\tau \rightarrow \mu \mu \mu$ or $\tau \rightarrow \mu \gamma$. The impact of improved lifetime measurements is illustrated in Fig.~\ref{fig:ildmassres}. In addition, b hadron decays modes with $\tau$ leptons are a crucial factor in the elucidation of flavour physics models.  Examples of b-hadron decay modes which access lepton universality tests are $B \rightarrow \tau \tau$ and $B \rightarrow \tau \nu$.  $B \rightarrow K^{*o}\tau^+\tau^-$ is an example of a particularly interesting electroweak-penguin mode, especially in the light of current anomalies~\cite{lhcbnature}.  The impact of the vertex detector performance for this channel is illustrated in Fig.~\ref{fig:leptonuniversality}.}
    \item{The dramatic expected improvement in precision on the measurement of electroweak observables which can be expected at FCC-ee relies on precision tracking and flavour tagging down to low angle acceptance.  This is particularly true of the b-quark electroweak measurements  $\mathrm{  A_{\rm FB}^b,0}$ and $ \mathrm{  R_b}$, where statistical improvements by a factor 800 and 2000 are expected~\cite{R1overview}}
    
    \item{The tracking system may also have an important role to play in particle identification, through dE/dx methods in the tracker, or with the addition of a tracking timing layer for time of flight measurements at the outside of the tracker (see also Section~\ref{pid}).}
\end{itemize}

\begin{figure}[htb]
    \centering
    \resizebox{0.9\textwidth}{!}{\includegraphics{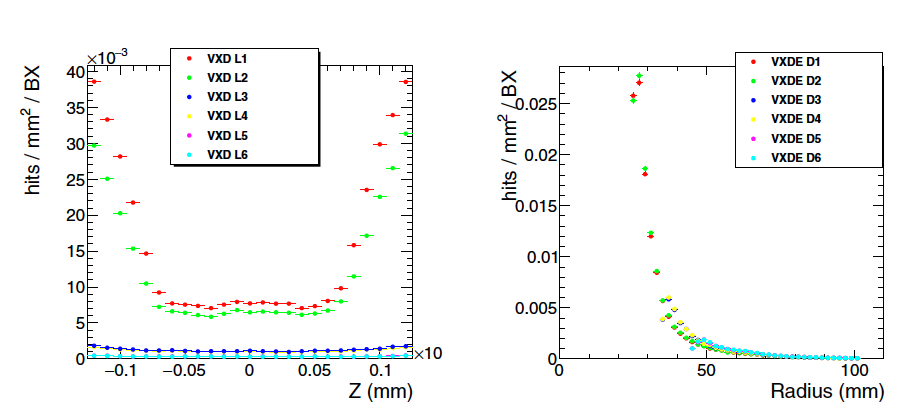}}
    \caption{Expected occupancy in the barrel and forward regions of the vertex detector, driven by incoherent pair creation~\cite{fcceecdr}.}
    \label{fig:occupancies}
\end{figure}

In order to address these challenges, the tracking system should satisfy the following requirements:

\begin{itemize}
    \item{{\bf Acceptance} \vspace{0.2cm}
    
    The FCC-ee will operate with low emittance beams colliding with a crossing angle of 30 mrad.  These parameters define the machine-detector interface, which covers a region  of 100\,mrad around the detector axis.  They also place a limit on the detector solenoid strength, which must be limited to 2\,T~\cite{machineinterface} to avoid unwanted beam emittance blow-up.  This has the consequence that the tracking volume must be larger than perhaps desirable, and the calorimeter may have to move outside a thin solenoid~\cite{Ilardi:2021ntv}.   The vertexing and tracking performance must be maintained across this full acceptance to exploit the full physics potential. In addition, the angular acceptance boundaries must be defined with great accuracy, of the order of 5-10\,$\mu$rad, for the high precision cross-section measurements. }\vspace{0.2cm}
    
    \item{{\bf Occupancy and Readout} \vspace{0.2cm}
    
    The detector readout and the front end pile up in the pixels must be able to cope with sustained physics rates of up to 100 kHz and backgrounds driven by synchrotron radiation and incoherent pair production.  At 365\,GeV operation, when the beams are separated by 994\,ns, the occupancies in the barrel vertex detector, illustrated in Fig.~\ref{fig:occupancies}, can reach 0.04 hits per mm$^2$ per bunch crossing at the innermost layer.  Taking into account an expected pixel pitch of approximately 25$\mu$m, a cluster multiplicity of 5 and a safety factor of 3 gives an occupancy of the vertex detector still below the level of  $10^{-3}$. Operating at the Z, the backgrounds are lower, however the bunch separation of 20\,ns combined with an expected detector time integration window of around 1\,$\mu$s yields similar occupancies~\cite{fcceecdr}.   Unlike detector designs for the ILC~\cite{Behnke:2013lya}, the operation cannot be in a power pulsed mode.  
    } \vspace{0.2cm}
    
    \item{{\bf Impact Parameter Resolution} \vspace{0.2cm}
    
    The target impact parameter resolution for individual tracks is {$5 \oplus {10} / (p_{\rm T} ~\rm{sin}^{\frac{1}{2}} \theta$)  $\mu {\rm m}$}, where $p_{\rm T}$ is the track transverse momentum in GeV/$c$.  Both the asymptotic and multiple scattering term in this formula are crucial for FCC-ee physics.  The transverse momentum of, for instance, muons from Z decays rely on the asymptotic term, whereas, tracks from, for instance the $K^*$ for the channel illustrated in Fig.~\ref{fig:leptonuniversality} have typical transverse momenta of 3-4\,GeV.  This resolution results in typical primary and secondary vertex resolutions (both transverse and longitudinal) of 3 and 7\,$\mu {\rm m}$ ~\cite{Monteil:2778908}. The system must be designed to ensure that the radial dimension can be calibrated to a relative precision of a few ppm, to ensure the same relative precision on e.g. the $\tau$ lepton lifetime and other weakly decaying particles.} 
    
    \vspace{0.2cm}
    
    \item{{\bf Momentum Resolution} \vspace{0.2cm}
    
    Excellent momentum resolution is required, with a target of $\Delta(1/(p_{\rm T})) \sim 2 \times 10^{-5} \oplus 1 \times 10^{-3}/(p_{\rm T} \sin \theta)$, driven by the requirements for precise recoil mass reconstruction, measurements of the Higgs mass, cross sections and branching ratios.} The possibility of constraining the possible point-to-point centre-of-mass energy uncertainties using the final state momentum distribution requires a high stability of the momentum scale~\cite{Blondel:2019jmp}; this will require a precise and continuous monitoring of both the tracker alignment and of the magnetic field. \vspace{0.2cm} 
    
    \item{{\bf Angular Resolution} \vspace{0.2cm}
    
    The typical muon angular resolution of the FCC-ee and other $e^+ e^-$ detectors, of the order of 0.1 mrad, is sufficient to have an impact of smaller than 1 MeV on the centre-of-mass energy determination, and can be measured with di-muon events with a more-than-adequate precision over the whole acceptance.~\cite{Blondel:2019jmp} } \vspace{0.2cm} 
    
    \item{{\bf Timing measurements}\vspace{0.2cm}
    
    Timing may be exploited in the tracking system to support PID, measurement of long lived particles, and to aid pattern reconstruction.  The vertex detector may be able to use the timing information to distinguish between early and late collisions from the beam bunches, exploiting the crossing angle. This would allow a check of beam-beam systematic uncertainty and a check on $\sqrt s$.  The possibility may exist to run a chromatisation scheme at the Higgs~\cite{telnov2020monochromatization} to scan a Higgs resonance within a single run.  For these kinds of physics goals the target track timing measurement would be of the order of 6 ps.} 
    
\end{itemize}

In order to achieve these goals the target material budget for the innermost layer of the vertex detector must be $0.2 \% X_0$ or better, with a target of less than $1\%$ for the whole vertex detector, and the individual hit resolution must be of the order of 3\,$\mu {\rm m}$.  The tracking power budget, which typically targets 40 mW/cm$^2$ or below, and cooling mechanism must be optimised to support the best possible transparency.  In addition, the tremendous statistics expected at FCC-ee will place unprecedented requirements on the stability, alignment, and calibration procedures for such a detector.  To give an example, the $\tau$ lifetime measurement, which will allow a precise test of lepton $\tau \mu$ universality, will be based on $10^{12}$ $\tau$ pairs and will reach an expected statistical precision of 0.001 fs, corresponding to a few tens of nanometers on the flight distance.  This will set stringent requirements on the offline alignment and overall radial scale of the vertex detector.


\section{Existing Detector Concepts}

\subsection{Tracking as implemented in the CLD concept}

\begin{figure}[htb]
    \centering
    \resizebox{0.4\textwidth}{!}{\includegraphics{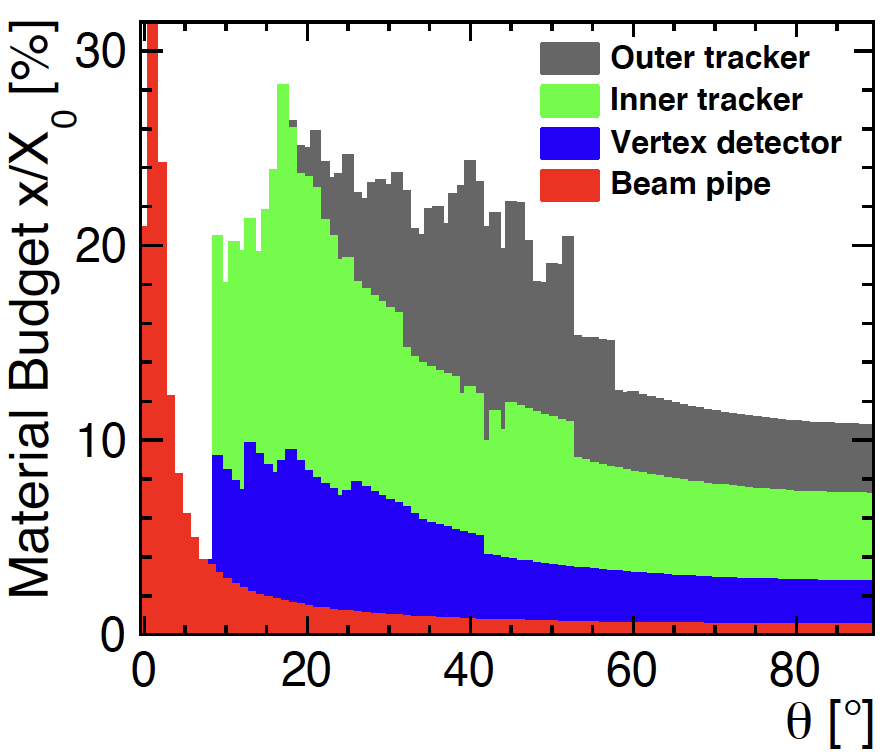}}
    \caption{Material budget distributions for the CLD detector concept}
    \label{fig:trackermaterialbudget}
\end{figure}

The CLD detector concept~\cite{Bacchetta:2019fmz} is an adaptation of a CLIC style detector design~\cite{Linssen:2012hp} to the FCC-ee environment, especially concerning the reduced value of the magnetic field which forces the tracking region to be extended from 1.5\,m to 2.1\,m. Another important change is the introduction of CO$_2$ evaporative cooling system as the pulsed power concept cannot be applied to FCC-ee. It is based on a full silicon concept with double layers of sensors on a common supporting carbon fibre structure for full coverage. It comprises 3 double layers of silicon pixel sensors and 3 double disks for the Vertex Detector, 3 double layers of short strips sensors and 7 forward double disks for the innermost tracking volume and a further 3 double layers and 4 double disks for the outer tracking volume.
The vertex detector features sensors with 25 $\times$ 25\,$\mu$m$^2$ pixels, an effective thickness of 50\,$\mu$m, an estimated total radiation length including cooling of 0.3$\%$ per single layer, and a total active area of about 0.35\,m$^2$. For the tracking region the proposed sensor is 200\,$\mu$m thick with 50\,$\mu$m $\times$ 1\,mm to 10\,mm long strips with the innermost double disk pixelated, similarly to the vertex detector. The estimated radiation length is 1$\%$ per layer for the region of the sensors, coolant and mechanical structure, and a further 2.5\% for the main cooling distribution pipes, main mechanical supports and cables. The total active area is about 196\,m$^2$. The total estimated tracker material budget is shown in Fig.~\ref{fig:trackermaterialbudget}.

\begin{figure}[htb]
    \centering
    \resizebox{0.4\textwidth}{!}{\includegraphics{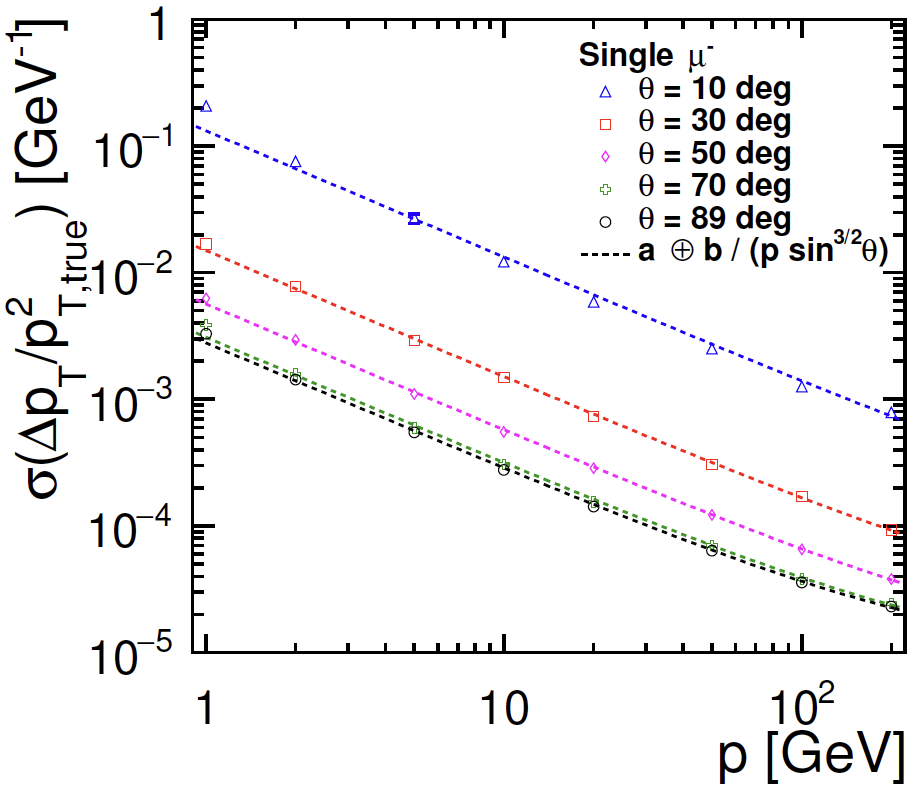}}
    \resizebox{0.4\textwidth}{!}{\includegraphics{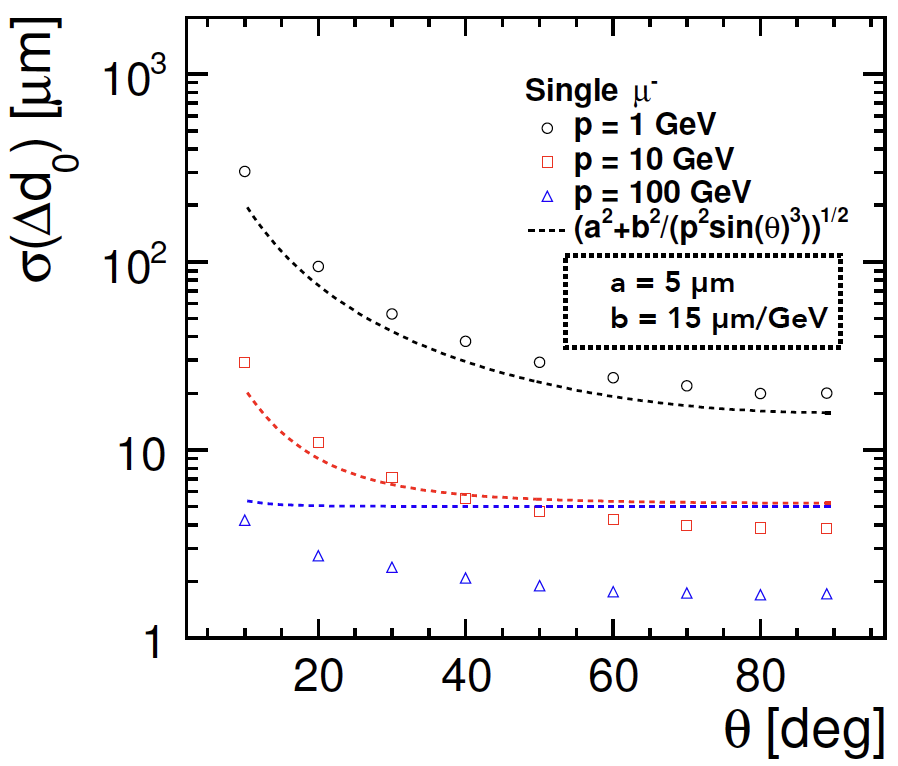}}
    \caption{Momentum resolution and impact parameter resolution achieved for the CLD detector concept}
    \label{fig:cldperformance}
\end{figure}

As an example of the overall performance, Fig.~\ref{fig:cldperformance} demonstrates a resolution better than $7 \times 10^{-5}$\,GeV$^{-1}$ for 45\,GeV muons at normal incidence corresponding to the required accuracy for the expected Z width measurement, and an impact-parameter resolution obtained for isolated muon tracks at various momenta achieving a resolution well below the high-momentum limit of 5\,$\mu$m at all polar angles.

\subsection{Tracking as implemented in the IDEA concept}

The current design of the IDEA detector concept~\cite{Antonello:2020tzq} features a silicon based vertex detector~\cite{Pancheri_2019} surrounded by a large drift chamber. The vertex detector currently considered is based on active monolithic pixel sensor technology relying on fully depleted high-resistivity substrates together with on-pixel sparsification and data-driven, time-stamped readout scheme. The target performance would be a resolution of a few microns with a total material of 0.15$\%$ - 0.3$\%$ X$_0$ per layer and power dissipation around 20 mW/cm$^2$ in order to avoid the need for active cooling.

A central aspect of the IDEA detector concept is a very light central drift chamber, which should achieve lower mass than the equivalent silicon-based tracking and provide better momentum resolution over the range of interest.  A novel feature of this detector is that adding timing information to the wires creates the possibility to count individual ionising events of the traversing track and dE/dx information.  This is called the cluster counting method and provides PID over most of the momentum range.

\begin{figure}[htb]
    \centering
    \resizebox{0.95\textwidth}{!}{\includegraphics{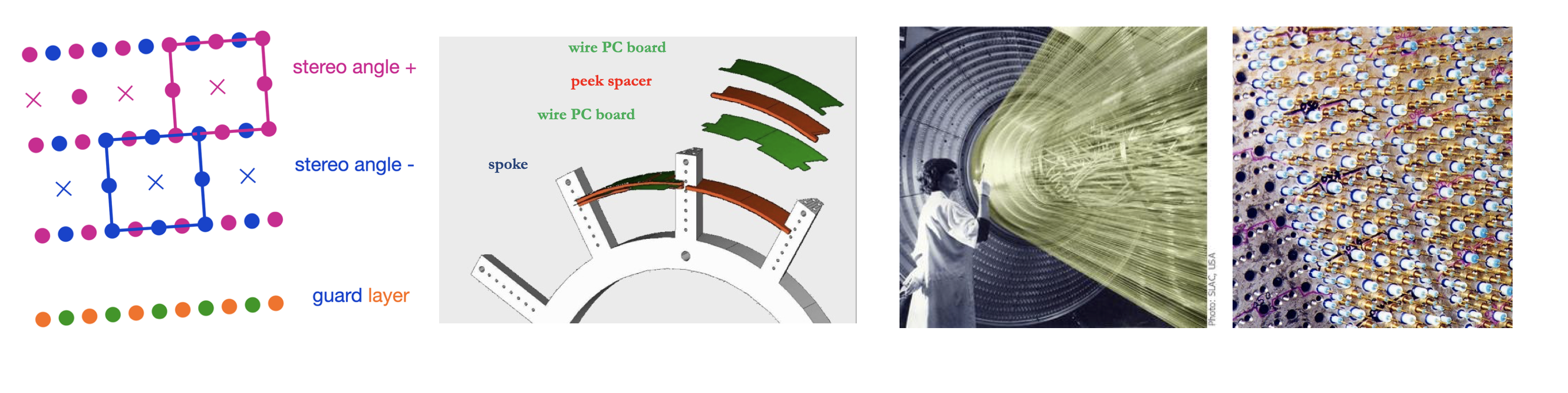}}
    \caption{Wiring concept for the IDEA tracker.  From left to right: Field/sense wire arrangement, mechanical mounting technique, photographs of achieved wiring for the MEGII drift chamber}
    \label{fig:ideawiring}
\end{figure}

The IDEA full stereo, high resolution, ultra-light drift chamber is inspired by the MEGII drift chamber concept~\cite{meg2} and its predecessors, such as the Mu2e I-tracker~\cite{Pezzullo:2018rso}.  The wires are laid out in a way which enmeshes the positive and negative stereo angle orientations, giving a high ratio of field to sense wires, and a high density of wires creating a more uniform equipotential surface.  There are almost 400k wires in total, requiring a non standard wiring procedure and a feed-through-less wiring system.  The wire support endplates also serve to contain the gas, allowing a reduction of material to approximation $10^{-3} X_0$ for the inner cylinder and a few times $10^{-2} X_0$ for the endplates.  The wiring technique is illustrated in Fig.~\ref{fig:ideawiring}, which shows how the wire PC board layers in green are built up on the frame, separated by precisely machined peek spacers.  The cluster counting exploits the fact that in He based gas mixtures the signals from each ionisation event are spread in time over a few ns.  With the help of a fast read-out electronics they can be efficiently identified and counted, giving a particle identification method with a better resolution than the integrated dE/dx method.

\begin{figure}[htb]
    \centering
    \resizebox{0.8\textwidth}{!}{\includegraphics{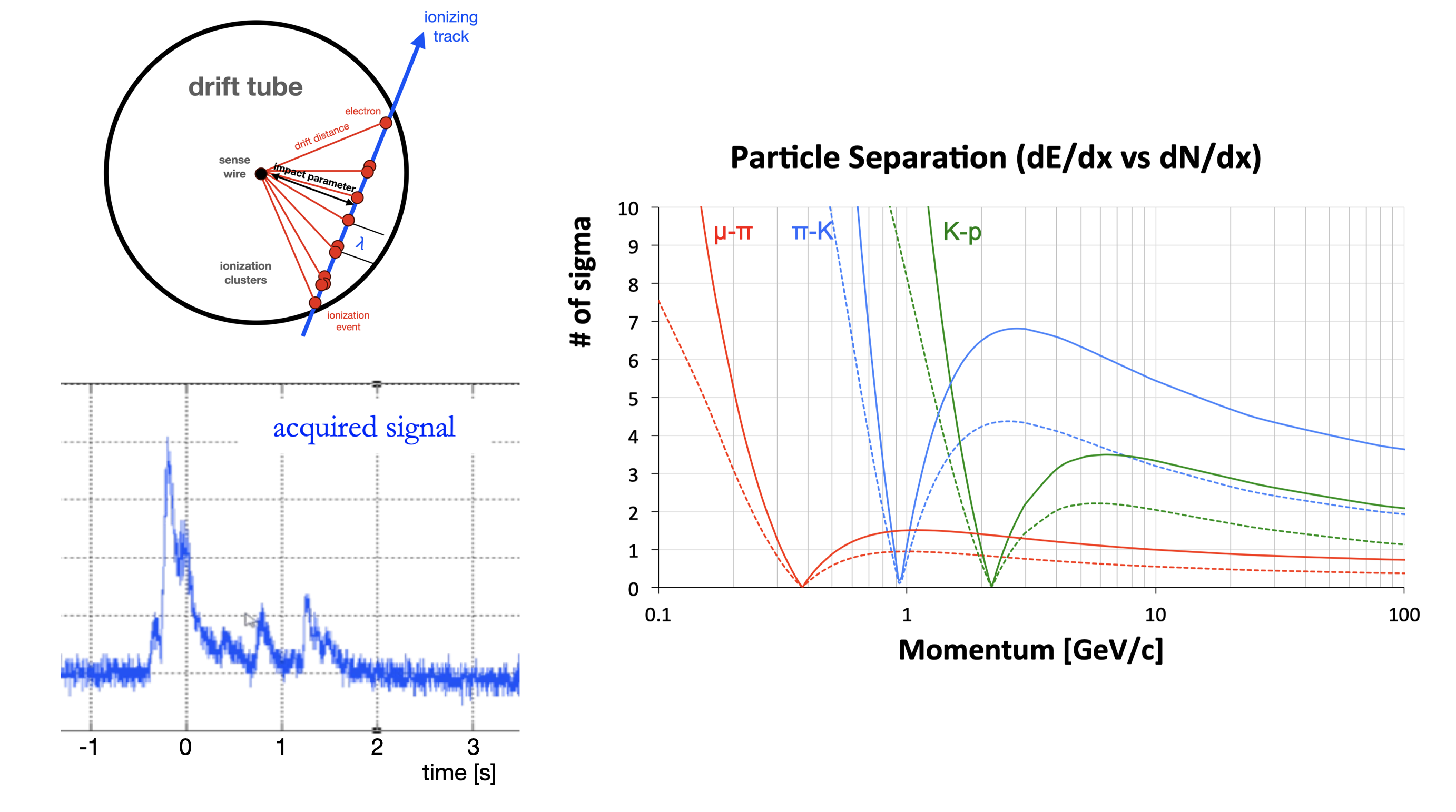}}
    \caption{Particle ID concept for the IDEA central tracker.  On the left, an illustration of the individual ionisation events and their distinct time of arrival as measured by the readout electronics. On the right, the expected particle separation performance (solid line), compared to a dE/dx approach (dashed line) for three different particle separation hypotheses.}
    \label{fig:taulifetime}
\end{figure}

\subsubsection{IDEA - Silicon wrapper}\label{pid}

The IDEA silicon wrapper offers a precise track 3D position measurement at the particles' exit point from the central drift chamber. It opens the possibility to provide an absolute reference for the calibration of the polar angle measurement, and hence define precisely the angular acceptance.  The wrapper encapsulates the tracker in both the barrel and forward regions.  In addition, if the silicon provides timing information this can can complement the PID range missing from IDEA by providing a  time of flight detector. For instance, to cover the pion/kaon loss of discimination around 1\,GeV, a timing of 0.5\,ns at 2\,m just outside the drift chamber would be sufficient to recover the performance.  An improved time resolution could strengthen the PID up to 5\,GeV.   Such a timing measurement will also be valuable for the reconstruction of secondary vertices in the search, or discovery, of massive long lived particles.


\section{Future Tracking Technologies}

In the rest of this document we focus on silicon based tracking devices and possible technological solutions which can be appropriate for FCC-ee implementation.  The IDEA concept described above is an example of a state of the art gaseous tracking solution, a full discussion of gaseous based alternatives lies outside the scope of this article.

\subsection{Silicon based tracking devices}

Silicon strips and pixels are currently the work horses of the vertexing and tracking programmes of the LHC experiments, and the current R\&D directions show excellent potential for addressing the needs of the vertex detector, optionally the tracker, and any external or timing layers of the tracking region of an FCC-ee detector.  Traditionally we distinguish two major categories of pixelated silicon tracking systems, hybrid and monolithic detectors.  In the hybrid system the sensor and front end chip are optimised separately and a fine pitch bump bonding technology is used to connect the sensor with the readout chip.  For monolithic devices the charge generation is integrated directly into the ASIC saving on the cost and complexity of the bump bonding step and allowing extremely thin sensors produced in a commercial process, suitable for the vertex region of an FCC detector. However, new technologies such as TSVs, microbumps, wafer stacking as well as alternative interconnection technologies blur the distinction between purely hybrid and monolithic approaches and will offer excellent potential for optimised devices for the future.  

\begin{figure}[htb]
    \centering
    \resizebox{0.95\textwidth}{!}{\includegraphics{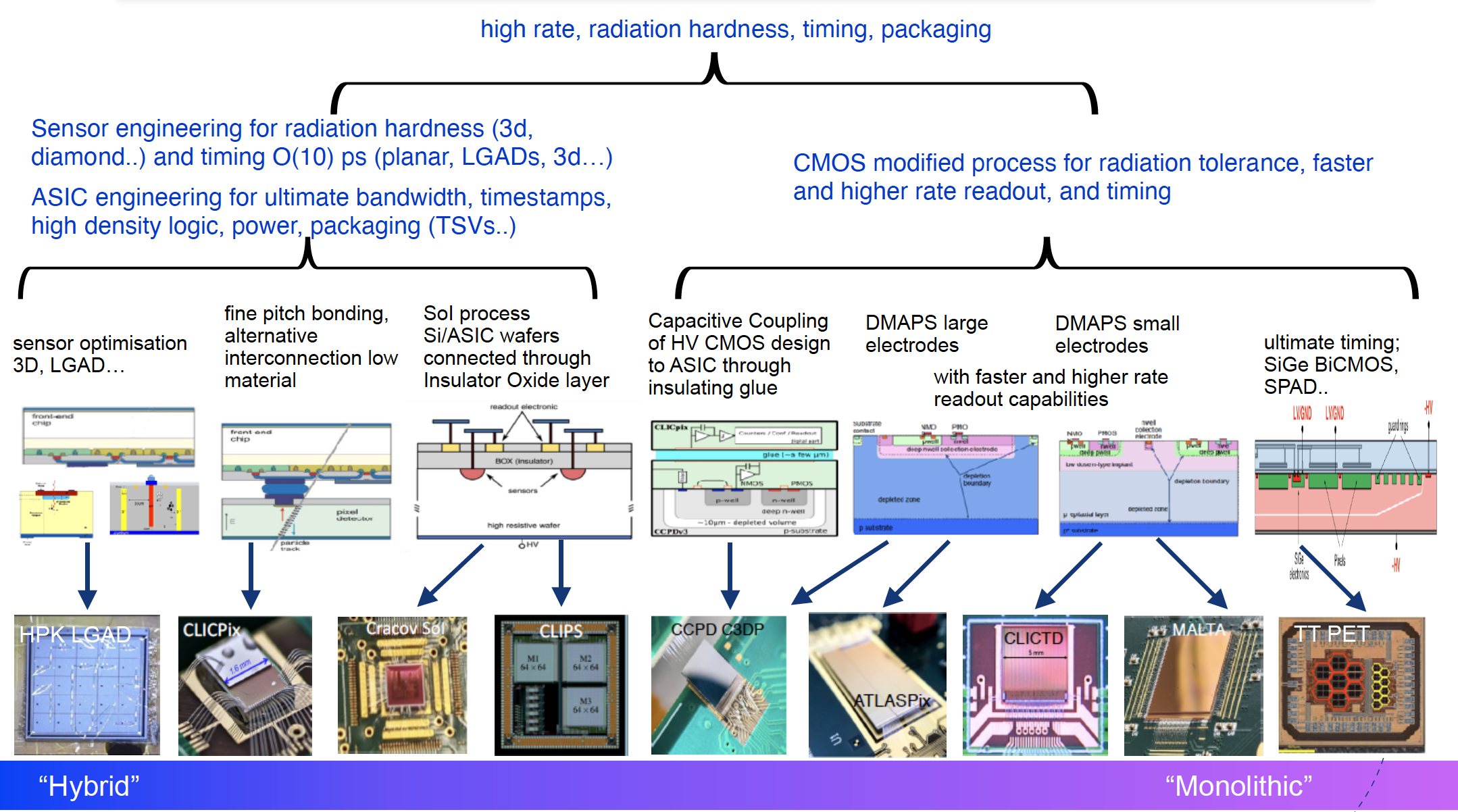}}
    \caption{Overview of R\&D developments for hybrid and monolithic pixel detectors}
    \label{fig:hybridmonolithic}
\end{figure}

Figure~\ref{fig:hybridmonolithic} presents an overview of several variants of hybrid and monolithic pixel developments.

\subsection{Monolithic CMOS MAPS}

The recent developments of low mass and low power CMOS MAPS (Monolithic Active Pixel Sensors) combined with the possibilities of large area coverage make this a very interesting technology option for FCC-ee. In recent years commercial CMOS processes with quadruple-well technologies to allow for full CMOS circuitry inside the pixel cell as well as high resistivity substrate wafers to enable depleting part or the full sensing volume have become available and are being explored by a large community.

The current state of the art in terms of installed detectors is represented by the upgrade of the ALICE ITS during LS2, which is the largest CMOS MAPS tracking detector ever built ($\approx$ 10 $\rm{m}^{2}$). This tracker has been installed in the ALICE experiment and is currently being commissioned.  The ALPIDE sensors are built in a commercial 180 nm imaging process and feature $27 \times 29 \mu{\rm m^2} $ pixels. The innermost three layers of this 7 layer tracker are constructed with a material budget of 0.35$\% X_0$ using 50\,$\mu$m thin MAPS connected to an Aluminium based flex cable and mounted on a low material budget mechanical support and cooling structure.  The ALPIDE use a high resistivity epitaxial layer as a sensing volume on top of a low resistivity p-type substrate.  By applying a moderate reverse bias ($\le$ 6V) the sensing volume around the collection electrode ($\approx$ 3\,$\mu$m diameter) can be partially depleted. This is fully compatible with operation in the ALICE radiation environment with total ionizing doses of $\approx 3$\,Mrad and NIEL fluences of $\approx$ 2 $\times$10$^{13}$ 1 MeV n$_{\rm eq}$ cm$^{2}$. During Pb-Pb collisions with an interaction rate of 50 kHz all events will be read out, while during p-p collisions the readout will be 400 kHz. The ALPIDE chip is a monolithic pixel chip with a small collection electrode, thus optimising the analogue power consumption. Together with a sparsified asynchronous readout without distribution of the clock to the matrix, power densities of about 300 mW/cm$^{2}$ for the innermost layers can be achieved, meeting the ALICE readout requirements.

HVCMOS sensors use commercial processes that embed NMOS and PMOS transistors in a single deep n-well that acts as a charge collection electrode~\cite{peric}. This allows to bias the substrate with a high negative voltage and to deplete the zone around the n-well. This technology has been chosen for the Mu3e experiment~\cite{mu3e} which has very stringent constraints on the material budget and requires the sensors to be thinned to 50 $\mu$m. The 2 $\times$ 2 cm$^{2}$ large sensors are mounted on a low mass service flexible printed circuit. The pixel detector is operated inside a dry helium atmosphere cooled by helium gas flow to further reduce multiple scattering~\cite{mu3e}.

\begin{figure}[htb]
    \centering
   \resizebox{0.5\textwidth}{!}{\includegraphics{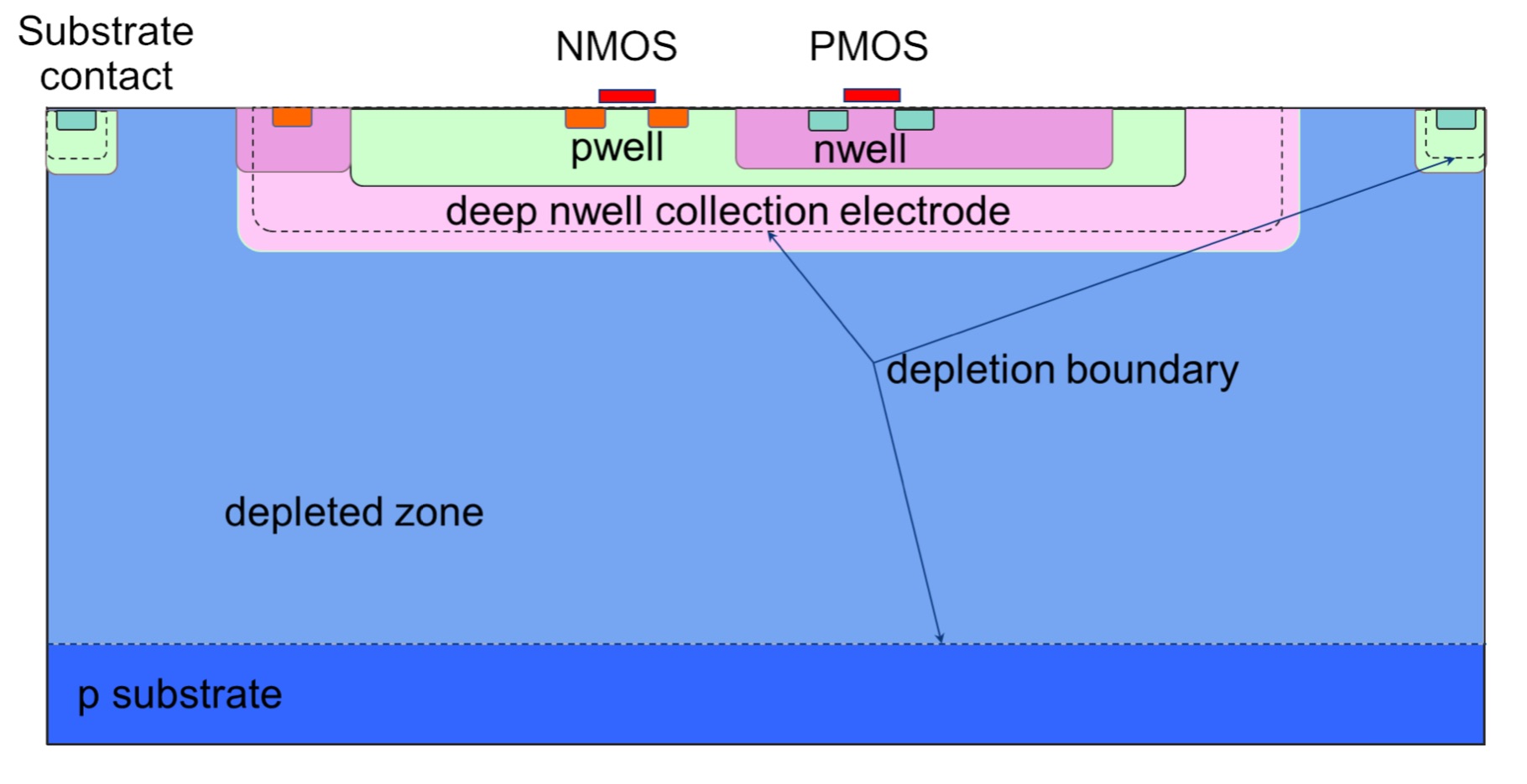}}
    \resizebox{0.45\textwidth}{!}{\includegraphics{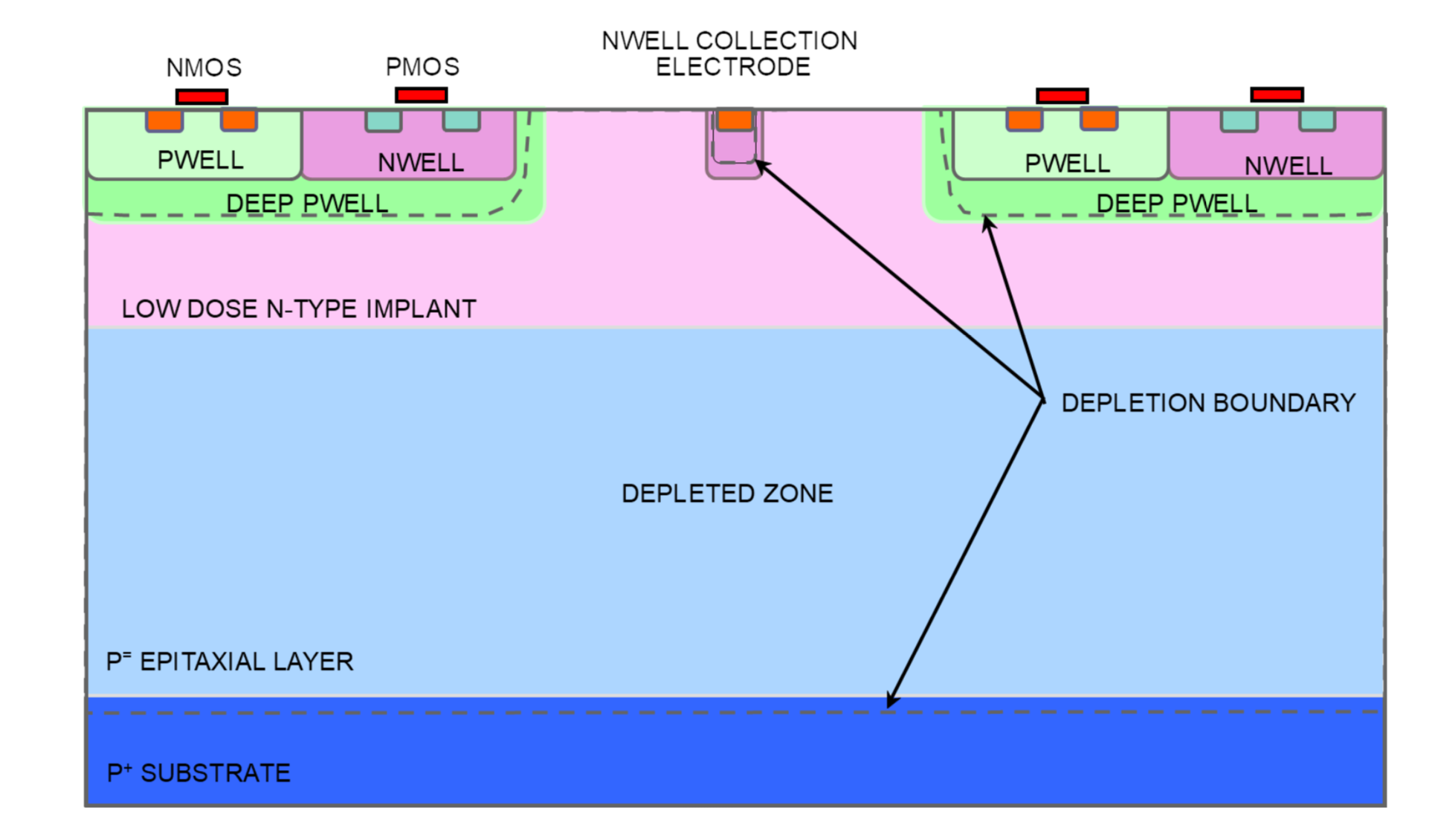}}
    \caption{Schematic cross sections of a pixel cell for large (left) and small (right) electrode design~\cite{Kugathasan:2019X9}}
    \label{fig:electrodes}
\end{figure}

The next generation monolithic sensors are moving in the direction of Depleted MAPS (DMAPS) driven by requirements for radiation hardness and faster readout as well as fast timing information. These sensors are built from high resistivity substrates and aim to deplete the sensing volume by applying few tens to several hundred volts reverse bias. By fully depleting the sensing volume the charges generated by a passing particle are collected by drift and not dominated by diffusion, which leads to longer charge collection times and potential charge trapping and loss in case of high radiation environments. Furthermore, fast charge collection by drift also improves the signal timing information. Several different design approaches are being studied for DMAPS, which can be coarsely divided into so called ‘‘small electrode" and ‘‘large electrode" designs, see also Fig.~\ref{fig:electrodes}. 

The small electrode approach with typical collection electrode diameters of a few microns seems especially promising, as it presents a much lower capacitance (a few fF as opposed to  values greater than 100fF), allows for implementation of smaller pixels in the same technology node, delivers faster signals, and a potential of better signal to noise and lower analogue power. This combined with a sparsified asynchronous readout scheme allows the reduction of the power density of the chip, however is dependent on the hit rate. Challenges of this approach include the routing of the signals to the chip periphery and will depend on the technology chosen and the feature size of it. Some of these concepts have already been incorporated in ASICs such as the MALTA/Monopix chip~\cite{SNOEYS201790} implemented in a modified Tower Jazz 180 nm process, and optimised for radiation hardness and speed~\cite{hsdt_pernegger}, and were also included in the CLICTD sensor chip~\cite{kremastiotis2020clictd}, optimised for extended efficiency for ultra thin silicon trackers.  

The MALTA chip is now being explored using high resistivity Cz silicon to enable high depletion depth~\cite{maltacz}, potentially using the full wafer thickness to increase the number of charges created by a passing particle while ensuring full depletion and fast charge collection by drift. These recent developments are proving that the area of sensor engineering, traditionally associated with hybrid pixel detectors, applies also to the monolithic domain. It opens up possibilities to engineer substrates and designs with improved charge collection properties and timing information.

Moving to a smaller feature size should allow smaller pitch as well as more room for routing and more functionality to be included in the pixel. Presently, several technology nodes with smaller feature sizes ~\cite{EPRD},~\cite{ecfaTF7}  are being explored. Adding further information on the charge deposited in the pixel cells and thus moving beyond a pure binary readout will help improving the hit resolution, while keeping in mind the available power budget. 



ALICE is already looking beyond the ITS for an ultra light inner barrel detector which could be installed in LS3.  This would consist of a new beam pipe with inner radius 16\,mm and sensing layers based on ultra thin, wafer scale sensors which can directly be curved to the required shape and operated with air cooling.   The application of such a self supported, curved circuit to FCC-ee would allow a decrease in average radius and would eliminate overlaps, acceptance loss and systematic effects due to varying radius or material from supports and services. It is made possible by the industrial availability of stitching which allows multi-reticle size ladders to be constructed of up to 30\,cm in 65 nm, together with the ability to thin the sensors to 20-40\,$\mu$m giving ultra thin and light detectors together with mechanical flexibility.  ALICE aims to exploit this process for a curved inner detector beyond LS3, and subsequently to construct an all silicon tracker for beyond LS4 operation consisting of 100\,${\rm m}^2$ double sided and curved layers. First test structures using stitching have been submitted for fabrication. While imaging sensors already use stitching to achieve large wafer scale sensors~\cite{ral},~\cite{ral2}, the application in HEP environments is now emerging and under study.

\begin{figure}[htb]
    \centering
    \resizebox{0.4\textwidth}{!}{\includegraphics{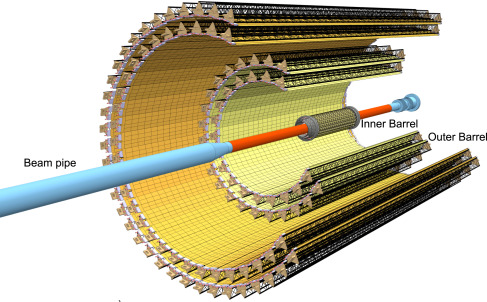}}
    \resizebox{0.4\textwidth}{!}{\includegraphics{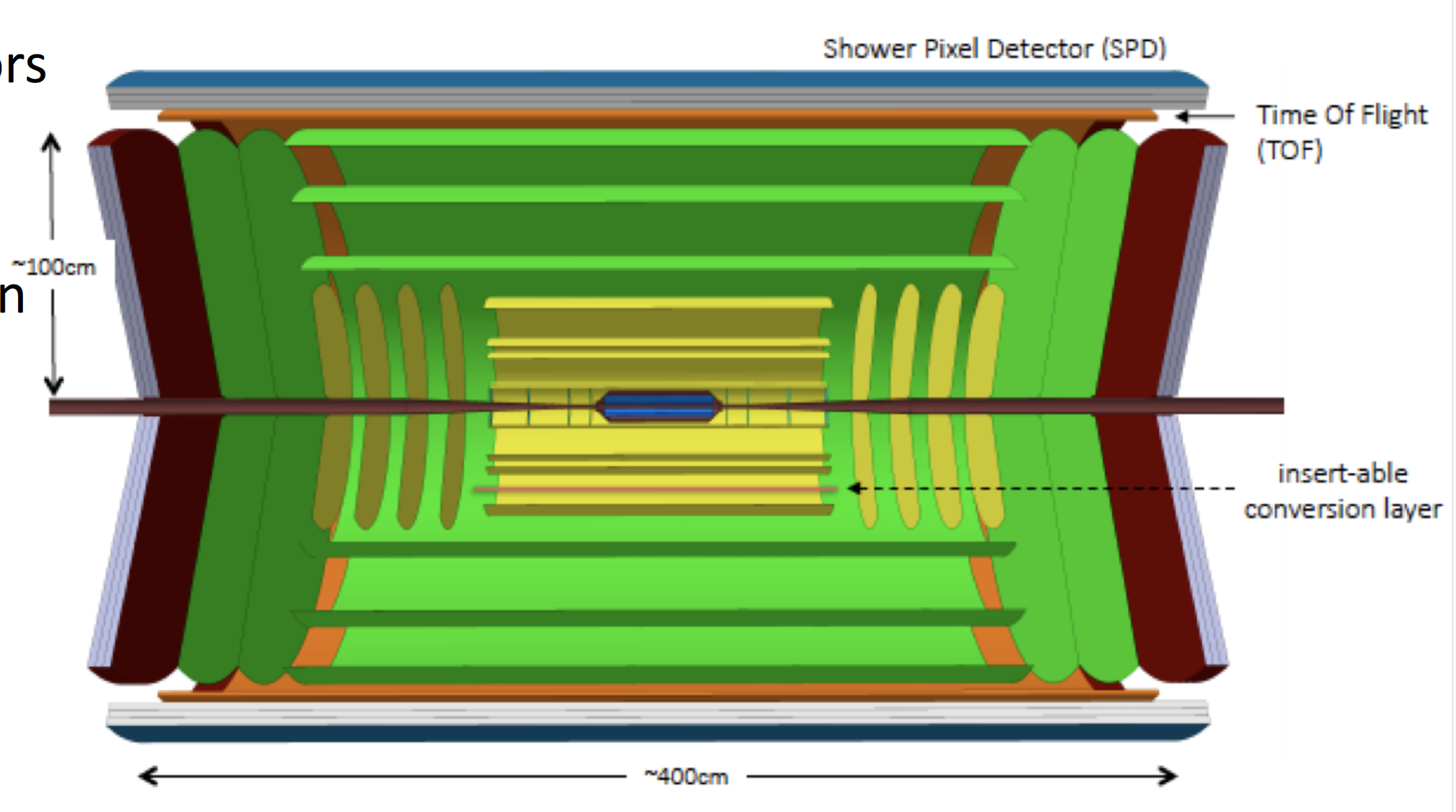}}
    \caption{Proposed exploitation of large area 65 nm MAPS devices, including truly curved stitched sensors for the inner region, for the ALICE LS3 and LS4 upgrades}
    \label{fig:aliceupgrades}
\end{figure}

\begin{figure}[htb]
    \centering
    \resizebox{0.95\textwidth}{!}{\includegraphics{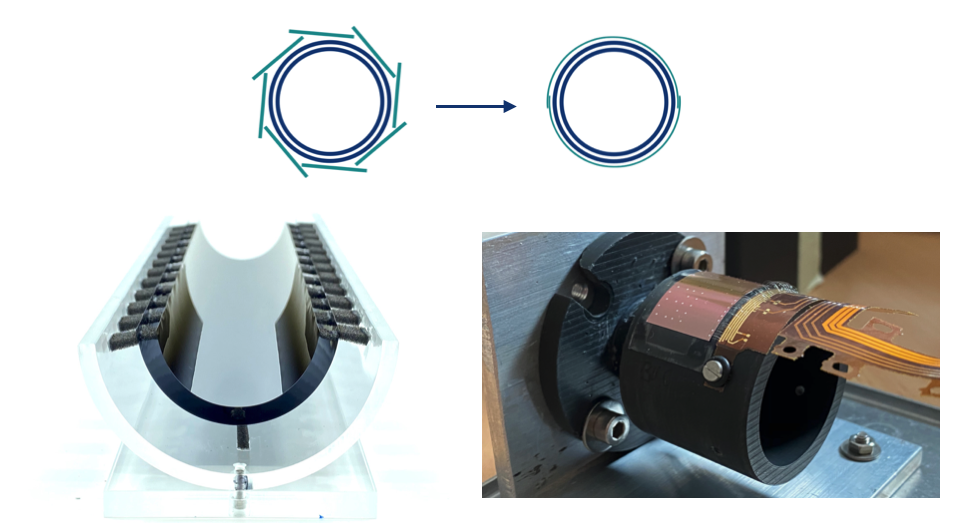}}
    \caption{The sketch illustrates the conceptual advantage of moving to large, curved sensors; the material contribution is lower and more uniform and the average radius decreases.  The bottom photographs~\cite{magnus2} show how this has been implemented for the ALPIDE R$\&$D; on the left, the mechanical integration and mounting of of thinned silicon wafers in the size of stitched sensors, on the right, a wire bonded curved ALPIDE sensor mounted on a rotating chuck.}
    \label{fig:alpidebonding}
\end{figure}

The challenges of such a design include the connection between the chips on the module level as well as off-module. The first connections to a curved sensor have been achieved successfully using aluminium wedge wirebonding, as shown in Fig.~\ref{fig:alpidebonding}. The curved sensor with bending radius of 1.8\,cm is mounted on a special jig for wire bonding; even at this radius the surface presented to the bonding machine for a given pad is essentially flat and the bonding can be completed successfully. To connect the sensor over the full width of 3\,cm, the jig has to be rotated to always present a flat bonding surface for the bonding head. 

\begin{figure}[htb]
    \centering
    \resizebox{0.3\textwidth}{!}{\includegraphics{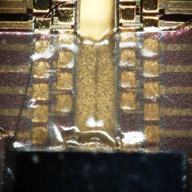}}
    \resizebox{0.4\textwidth}{!}{\includegraphics{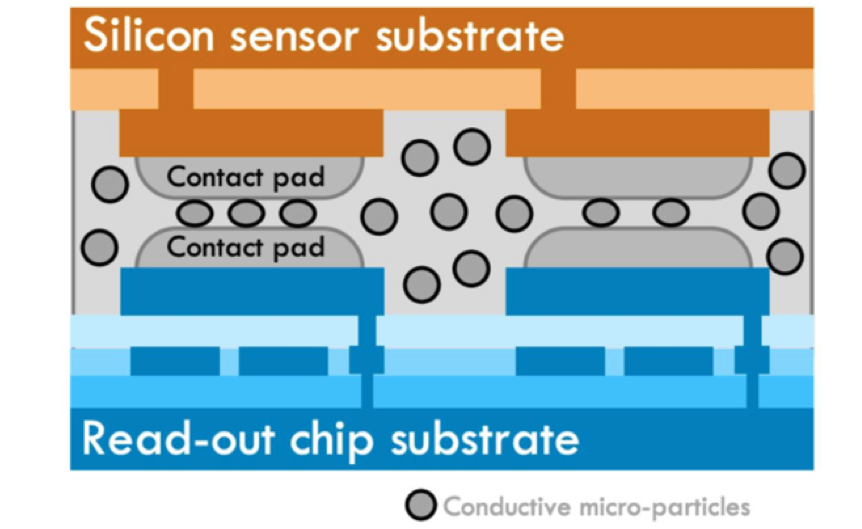}}
    \caption{Two MALTA chips with ACF~\cite{DTannual20} (left) and schematic view of ACF connected components~\cite{ACF} (right).}
    \label{fig:MALTA_ACF}
\end{figure}

The study of alternative interconnection technologies beyond the traditionally used aluminium wedge wire bonding focuses on building large area modules with reduced and optimised interconnection schemes~\cite{EPRD}. This includes data and power transfer from chip to chip, as implemented in the MALTA chip, as well as the study of connection off-module. ACF (Anisotropic Conductive Film) is here a promising candidate to achieve a low mass, simplified connected between chip pads and a flexible printed circuit. Figure~\ref{fig:MALTA_ACF} shows ACF film deposited on two MALTA chips before flip chip connecting a silicon bridge that connects the data and power pads from one chip with the neighbouring chip~\cite{ACF}. The data accumulated in one chip will be transferred via the bridge contact to the neighbouring chip, merged with the data of the second chip and read out. This concept it presently under study to be extended to four MALTA chips, where the data from all chips are read out only via the last chip in the chain. This technique can also be used for serial powering and thus reducing the material budget for services. Due to the pad size of 88\,$\mu$m side length the connections between chips is presently carried out using either the silicon bridge or aluminium wedge wire bonding~\cite{modules}. 

\begin{figure}[htb]
    \centering
    \resizebox{0.8\textwidth}{!}{\includegraphics{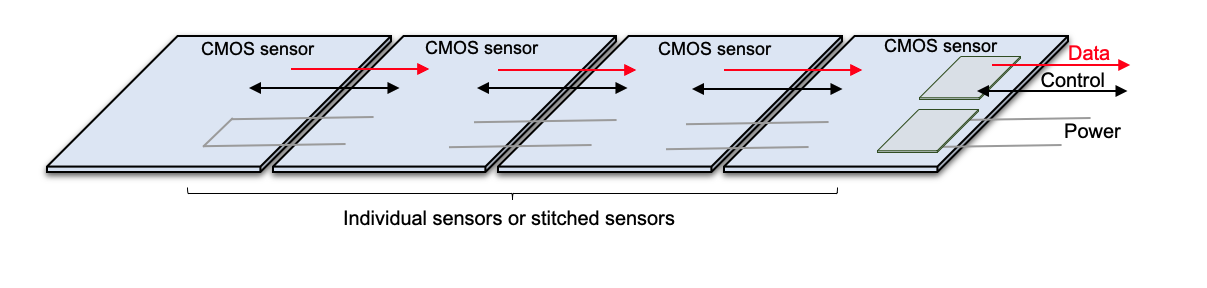}}
    \caption{Schematic sketch of a light weight and large area module based on CMOS sensors~\cite{riedler_ecfatf3}.}
    \label{fig:future_module}
\end{figure}

Further studies, such as the use of RDLs (ReDistribution Layers) will allow the scaling of the chip pad connectivity with the typical size of flexible printed circuit pads, so that assembly techniques can be simplified for larger module numbers and thus larger surfaces, see schematic sketch in Fig.~\ref{fig:future_module}. 

\subsection{DEPFET, FPCCD technologies}

The DEPFET and FPCCD technologies are examples of sensors with a split charge amplification and readout scheme.  Each pixel is a p channel FET on a fully depleted bulk; the electrons are stored in the internal gate, and the accumulated charge is removed after readout by a clear contact.  After processing the wafer backside may have deep areas removed via anisotropic deep etching - the sensors are supported by the monolithically integrated silicon frame and the thickness of the sensitive area is almost a free parameter.  This has allowed the DEPFET pixel detector at Belle II to be constructed with a record 0.2$\%$ $X_0$ per layer.  The next generation of DEPFET detectors will further optimise the shape of the drain implants to improve the signal, reduce the power consumption, and improve the radiation hardness~\cite{BORONAT2016982}, and in a similar way to the MAPS developments, curved sensors might be employed.

FPCCDs offer the the possibility to achieve very small pixels, $\approx$ 5 $\mu$m square, with a sensitive layer that is fully depleted. The signal charge is transferred over a longer distance, up to a few cm, which can lead to radiation induced charge transfer inefficiencies. While radiation damage induced defects cause charge loss by trapping, several techniques to compensate this effect are under study. Using so called fat-zero charge injection methods which fill the radiation induced lattice defects are being investigated. Furthermore, notch-CDD designs that optimise the charge transfer channel width from one pixel to another as well as operation at low temperature are employed to reduce the radiation induced charge transfer inefficiencies~\cite{fpccd}.

\subsection{3D Integration, SoI}

Industrial developments are strongly focused on heterogenous integration technologies which will allow further reductions in cost and power over the coming years. Several of these technologies, such as 3D stacking are also been studied for future HEP applications\cite{ecfa_roadmap,pristauz}. The developments in this area are dependent on availability of processes for R\&D activities, but are a very promising path for future developments which will increase the functionality and performance of silicon trackers.

\begin{figure}[htb]
    \centering
    \resizebox{0.6\textwidth}{!}{\includegraphics{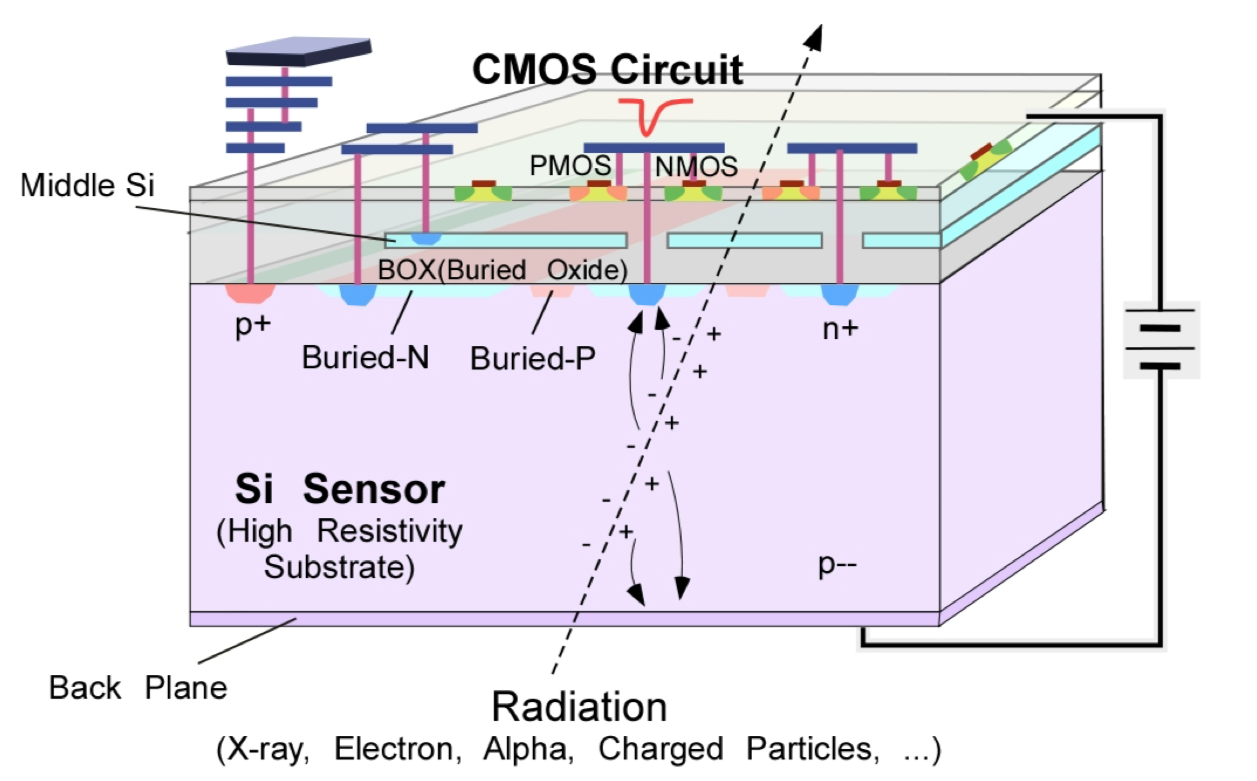}}
    \caption{Schematic view of a SOI structure~\cite{Arai:2017dni}.}
    \label{fig:soi}
\end{figure}

The developments of SOI (Silicon On Insulator) detectors present a two-tiered separation of the readout electronics and the sensing layer~\cite{Tsuboyama:2019gzj}. An SOI wafer with a high resistivity sensor part is connected to a thin CMOS chip using oxide bonding. Figure~\ref{fig:soi} shows a schematic SOI structure indicating the different tiers and signal generation by a passing particle. This approach has been further developed using buried wells, double SOI and pinned depleted diodes to address signal cross talk and radiation tolerance~\cite{Arai:2017dni}. 

\subsection{Solutions based on hybrid pixel sensors}

Hybrid pixel solutions, which make up the majority of currently installed systems, offer the possibility to separately optimise the sensor characteristics while being able to benefit from standard commercial processes for a high degree of functionality for the ASIC.  They are typically able to cope with very high rates and high radiation environments.  Challenges include reducing the material for the sensor, ASIC and interconnection region, and the costs and technological challenges of fine pitch bump bonding.  The hybrid pixel solution is of particular interest for FCC-ee due to the possibility of adding sensor and ASIC timing capability, which could be exploited in external layers or in the wrapper layer to support PID capability.  A candidate sensor technology for fast time stamp applications is that of LGADs (Low-Gain Avalanche Diodes~\cite{hartmut}) which incorporate a thin multiplication layer which aims to supply a small gain which is sufficient to give a reliable timing signal.  Such a technology has been shown to be capable of delivering 20~ps resolution or better, depending on the signal size and sensor thickness~\cite{KRAMBERGER201926}.  The concept of a timing layer using LGADs is being exploited for both the ATLAS and CMS upgrade timing layers in the endcap regions, which aim for a system level timing resolution of 30~ps with 10-15 m$^2$ of silicon, for installation in 2023-2025~\cite{Butler:2055167,CERN-LHCC-2015-020,Mazza:2019dkn,Lazarovits:2020nfz}.  The LGAD technology is being improved to allow a fine pitch pixel readout, which in traditional LGAD designs is impeded by the no-gain region between the pixels.  This can be addressed with solutions such as using trenches for pad isolation, the so-called TI-LGAD~\cite{paternoster}, moving the gain region to the other side of the sensor (``iLGAD - inverted LGADs~\cite{cartiglia_ilgad}") or implementing a resistive readout solution, the so-called AC-LGAD or RSD, which can achieve excellent position resolution even with large pixels, freeing up space in the ASIC to allow timing functionality to be added~\cite{tomago}.  
An alternative approach could be monolithic BiCMOS, exploiting the properties of SiGe transistors, where the development is now beginning, and where the timing performance could also be enhanced with the use of internal gain.

\begin{figure}[htb]
    \centering
    \resizebox{0.55\textwidth}{!}{\includegraphics{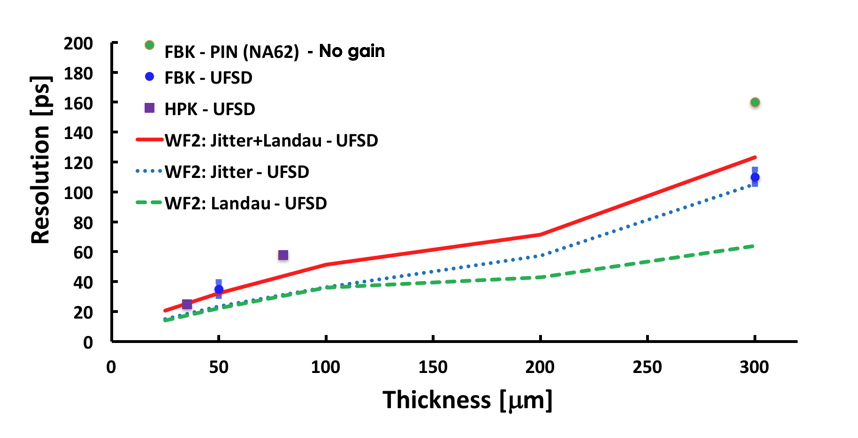}}
    \resizebox{0.3\textwidth}{!}{\includegraphics{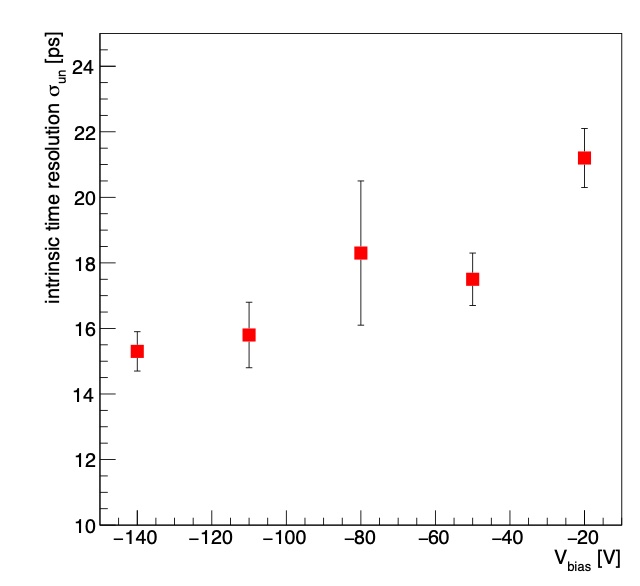}}
    \caption{Left: Comparison between simulation and a range of measurements for LGAD sensors~\cite{hartmut}.  Typically time resolutions of down to 25 ps are achieveable with 50\,$\mu$m square pixels for detector thicknesses of 45-55\,$\mu$m Right: Recent time resolution results from trench optimised 3D sensors.  Subtracting the electronics contribution a time resolution of $\-$10-ps is estimated for the 55\,$\mu$m square pixels at 140V bias.~\cite{Anderlini_2020}}
    \label{fig:hybrid_timing}
\end{figure}

A further possibility for timing detectors which do not incorporate internal gain is the use of silicon sensors with a three dimensional (3-D) architecture~\cite{PARKER1997328}, where the n and p electrodes penetrate fully or partially through the silicon substrate.  This is a rapidly evolving technology which was first successfully deployed at the LHC for the ATLAS IBL~\cite{DaVia:2012ay} and is currently being evaluated for the Phase II LHC Upgrades~\cite{SZUMLAK2020162187}.  Due to the inherently short drift distance, independent of the sensor thickness, the 3D sensors can be very fast, and have good radiation tolerance, and low operation voltages and power dissipation.  One of the challenges for timing resolution comes from the variations in the weighting field due to the hit position, and the fill factor, which may be significant for small cell sizes and certain track inclinations.  One way to tackle the weighting field variations is with a trench design, as developed by the TIMESPOT collaboration, and for which intrinsic detector resolutions of the order of 10 ps have been demonstrated in testbeam, as shown in Fig.~\ref{fig:hybrid_timing} (right).  An implementation of timing sensors in a wrapper layer for FCC-ee, with relaxed requirements on pitch may be able to benefit from the enhanced time resolution of traditional column 3D sensors in a configuration with multiple cells connected together~\cite{kramberger_timing}.

The development of any sensor with timing capability must be done hand in hand with the accompanying IC technology.  Applications at FCC-ee can expect to benefit from the current ongoing R$\&$D into ASIC developments for HEP applications including LHC Upgrades, as well as imaging applications.  Recent developments include the Timepix4 ASIC~\cite{timepix4}, a full size 4-side tileable chip with high rate imaging capabilities which has been successfully developed in 65\,nm~CMOS technology and provides a time stamp binning resolution of 195~ps (RMS 56~ps).  The Timespot demonstrator ASIC~\cite{TimeSpOT:2020gln}, is being developed in 28\,nm CMOS technology and aims for ultimate time resolution using a CSA inverter input stage.   The LHCb experiment plans a Phase II Upgrade~\cite{lhcbupgrade2}, due for installation after LS4, which will require a high granularity pixel vertex detector capable of hit timestamps, operating at high speed in a high radiation environment.  Such developments provide a promising path forward for an eventual hybrid pixel layer implementation at FCC-ee.

\subsection{Mechanical Integration and Interconnection Technologies}

\subsubsection{Mechanics and Cooling}

The Cooling and mechanical design of a FCC-ee tracker poses many challenges.  In order to benefit from progress towards very thin sensors, the supports and services must also reduce the amount of mass and the detector must benefit from an integrated design.  The very high luminosities and $\rm e^+ e^-$ cross section at the Z peak imply relative experimental systematic uncertainties at the $10^{-5}$ level, required to match the statistical accuracies.  Hence mechanical stability is also a crucial consideration, whether deviations come from internal effects such as vibrations or thermal effects, or external stimuli such as cavern floor movements, earthquakes or stress from other subsystems.

\begin{figure}[htb]
    \centering
   \resizebox{0.45\textwidth}{!}{\includegraphics{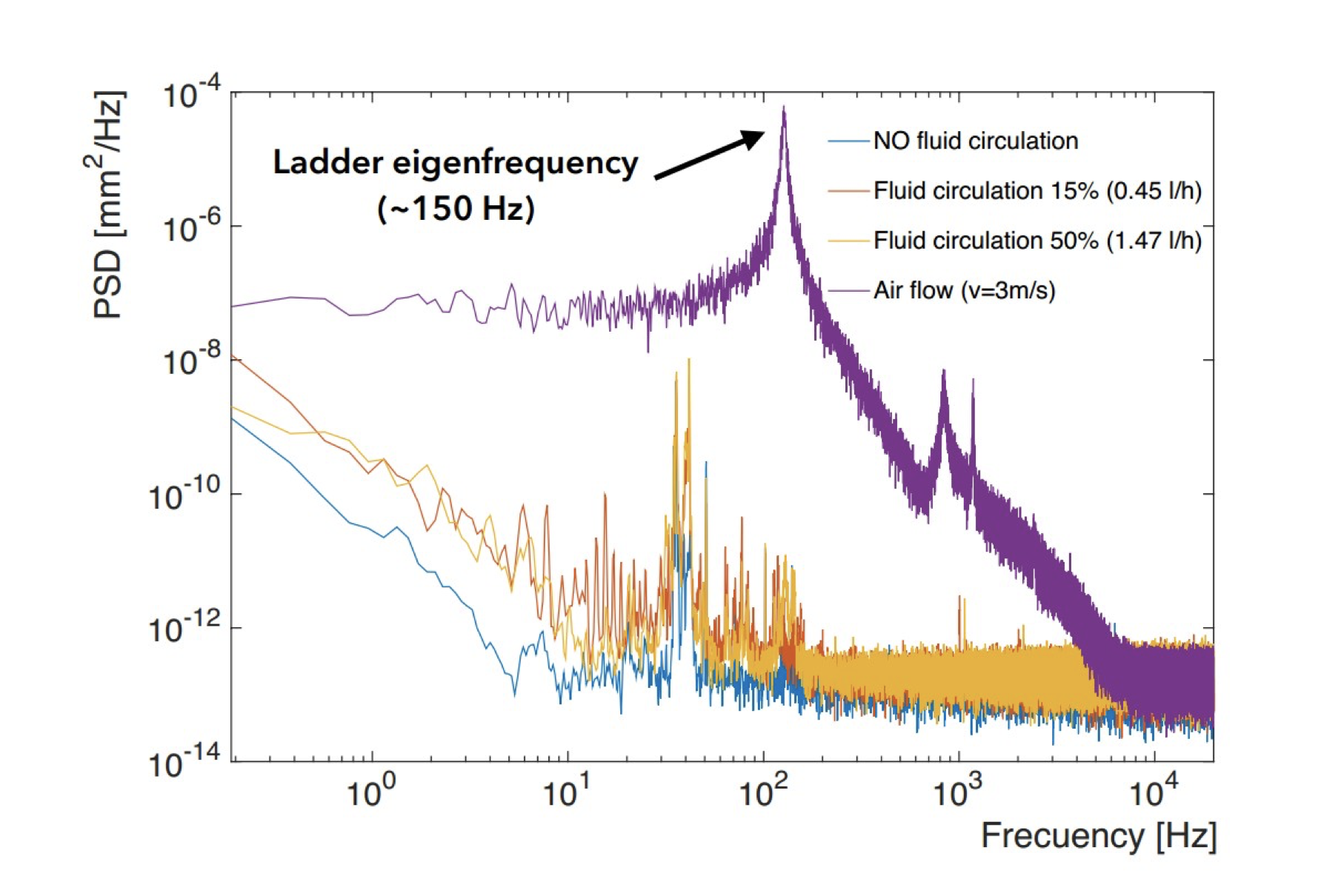}}    \resizebox{0.45\textwidth}{!}{\includegraphics{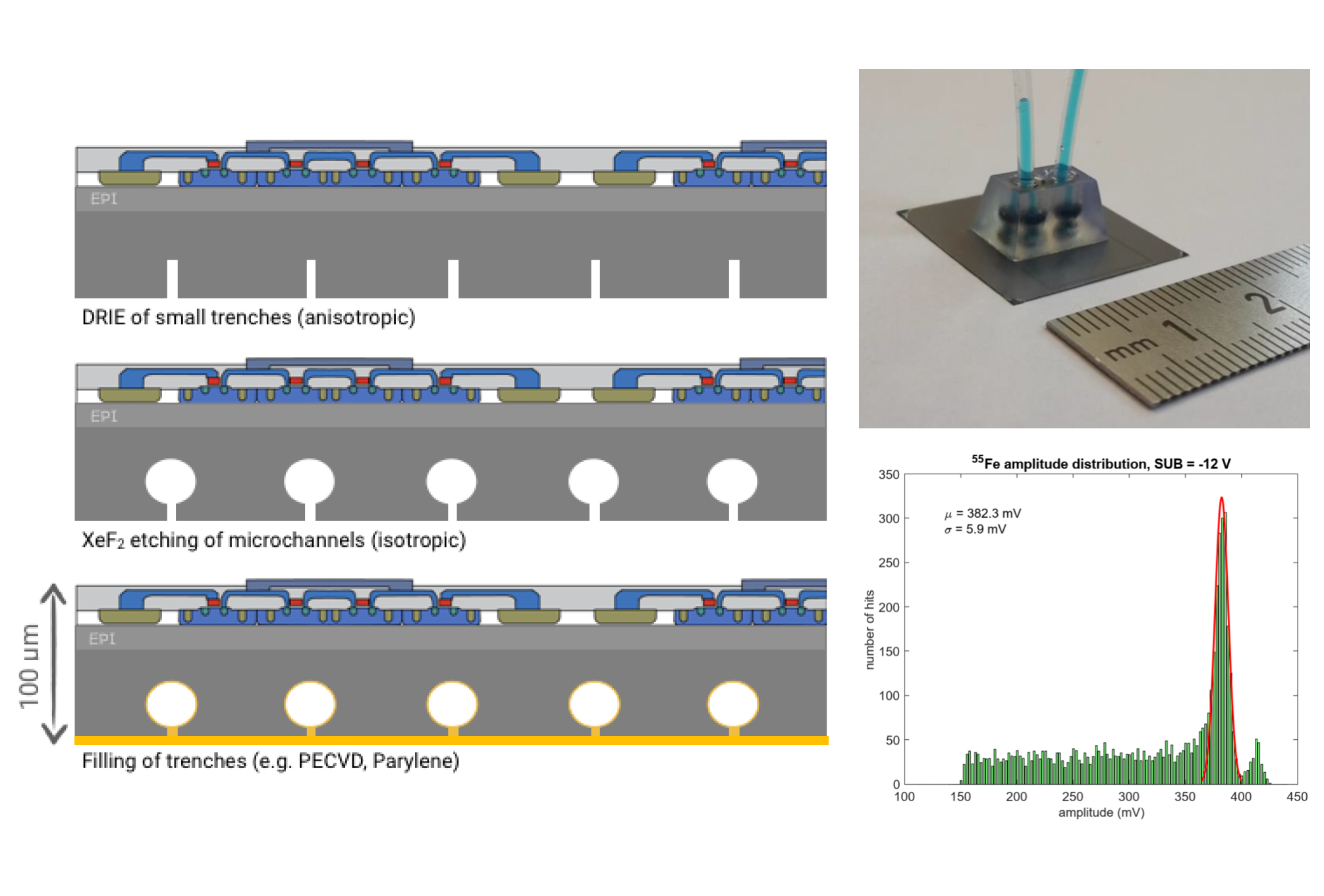}}
    \caption{Left: Study of the relative importance of different types of cooling for the stability of silicon ladders ~\cite{Andricek_2016} Right: Example of the technique to integrate cooling channels directly in the substrate of an active sensor, with a photograph of a fully functional demonstrator MALTA chip and the Fe$_{55}$ source scan showing that the sensor is fully functional}
    \label{fig:mechvib}
\end{figure}

The global design of the FCC-ee tracker will need to take advantage of the lightweight next generation solutions for the mechanical supports and cooling. In the ideal case the detector active cooling circuits are completely removed and the heat is removed by a forced airflow. In this case attention must be paid to the vibrations which can be induced in thin silicon ladders.  The design must be backed with simulations and realistic mechanical measurements~\cite{Viehhauser_2015}.  An example of such a study is shown in Fig.~\ref{fig:mechvib}. This approach is currently under investigation for the ALICE ITS3~\cite{alicetdr,magnus}, where the stitched, curved sensors are supported by ultra-light carbon foam supports which also act as radiators, and the whole device is contained within an external carbon exoskeleton.  The engineering model is currently being tested in a wind tunnel for thermal and mechanical stability.  It may be that air cooling cannot be implemented, if for instance there are regions with more dense power consumption, or if the size and complexity of the detector does not allow large quantities of air to be introduced in well controlled temperature conditions.  If a small increase in material budget is acceptable, microchannel cooling~\cite{lhcbuchan} represents one very promising area of development.  Such a solution has already been implemented for the LHCb pixel detector upgrade, where the coolant is evaporative CO$_2$ which circulates in tiny microchannels embedded within a cooling plate consisting of a silicon wafer.  Such a system has the advantage of a low and uniform mass distribution, a high thermal efficiency, a good CTE match between substrate and sensor, and a flexible geometry allowing the coolant to be brought precisely below the needed regions.    A low material contribution is possible, for instance in the case of the GTK microchannel cooling plates of the NA62 experiment~\cite{na62uchan}, which are thinned in the acceptance to a final contribution of 0.13\% X$_0$.  Recent technological advances for the silicon etched microchannel solution allow the cooling channels to be directly integrated in the sensor substrate.  This has been demonstrated for instance in the buried channel process developed by FBK which has been demonstrated on a single MALTA die to be compatible with the CMOS processing~\cite{Mapellivci} and showing full functionality. Similarly, integrated micro-channel cooling to DEPFET detectors is also under consideration~\cite{Vos:2017uin}.   For future microchannel applications a variety of processes in additive manufacturing are being considered.  These allow in principle a large choice of materials such as metals, ceramics or acrylics, and have the advantage great flexibility in the geometric forms, as well as being able to address the issue of the interconnections, where for the etched silicon solution the connections tend to be more bulky and fragile.  A similar approach is taken by the concept of a microvascular network embedded in a carbon cold-plate~\cite{carbonvascular}, which could be adapted to silicon detector cooling supports.  

\begin{figure}[htb]
    \centering
   \resizebox{0.75\textwidth}{!}{\includegraphics{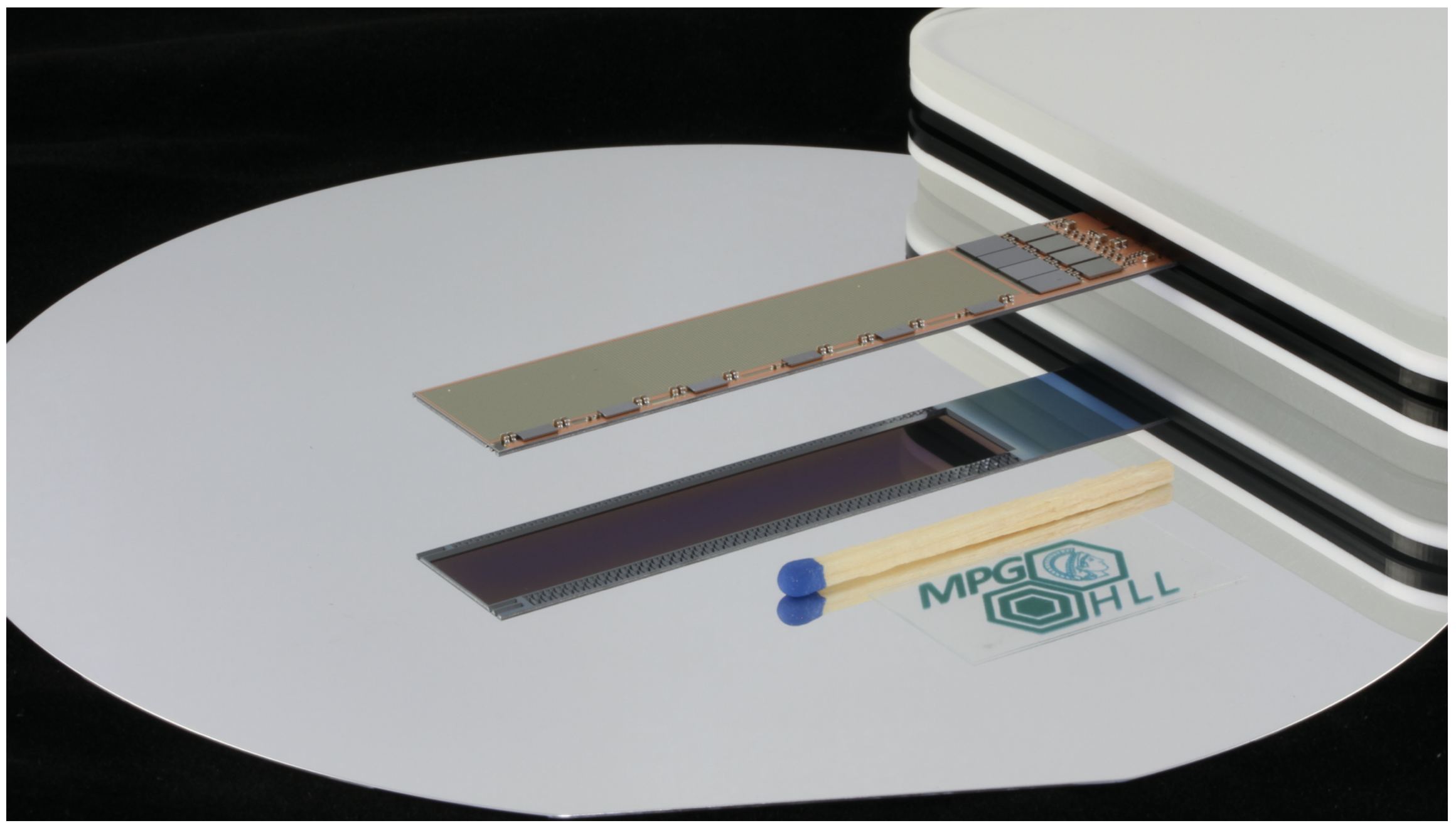}}    
    \caption{DEPFET ladder for BELLEII, illustrating integration of supports, routing and electronics within one all-silicon layer}
    \label{fig:depfet_integration}
\end{figure}

An excellent example of the integration of supports, routing and electronics in one thin all-silicon ladder detector is given by the DEPFETs of Belle II, as illustrated in Fig.~\ref{fig:depfet_integration}.  Whichever technological solution is chosen for a future FCC-ee detector the way in which the sensors are integrated into modules will be critical due to the need to maintain low material throughout the tracking region.  The FCC-ee vertex detector will rely on enabling technologies for module packaging, including thinning and dicing techniques, fine pitch bump bonding, or design of multichip modules incorporating concepts such as serial powering.

\section{Conclusions}
\label{section:conclusion}

The tracking system of a future FCC-ee detector will require innovative technological solutions for the sensors, mechanics and readout in order to address the needs for high precision and low mass. New solutions have to be found be found for alignment and field stability, so that tracker systematics match the new level of statistical accuracy.  The chosen solutions must take advantage of current trends in cutting edge technology.  In particular, advances in low mass, low power and fine pixel monolithic sensors may be applicable in the inner most layers, sensors with timing capability may be used in outer layers or as an outer wrapper, and the main tracking body may employ for example an all silicon solution or a low mass drift chamber.  The detector will act as the interface to the machine and the design and constraints on the beampipe design must be closely integrated with the evolution of the detector layout.  
%
%
%

\bibliographystyle{jhep}
\bibliography{references}
\end{document}